\documentclass{appolb}
\usepackage{graphicx}
\usepackage{url}
\usepackage{hyperref}
\usepackage[super,sort&compress,comma]{natbib}

\begin{document}
\bibliographystyle{Science}

\title{Fundamental physics at the strangeness frontier at DA$\Phi$NE. \newline Outline of a proposal for future measurements. }

\author{C. Curceanu, C. Guaraldo, A. Scordo, D. Sirghi, 
\address{Laboratori Nazionali di Frascati INFN, Via E. Fermi 54, Frascati, Italy}\vspace{0.5cm} 
\newline {K. Piscicchia}
\address{Museo Storico della Fisica e Centro Studi e Ricerche Enrico Fermi, Rome, Italy}\vspace{0.5cm}
\newline {C. Amsler, J. Zmeskal}
\address{Stefan Meyer Institute of the Austrian Academy of Sciences (SMI), Wien, Austria}\vspace{0.5cm}
\newline {D. Bosnar}
\address{Department of Physics, Faculty of Science, University of Zagreb, Zagreb, Croatia}\vspace{0.5cm}
\newline {S. Eidelman}
\address{Budker Institute of Nuclear Physics (SB RAS), Novosibirsk and Lebedev Physical Institute (RAS), Moscow, Russia}\vspace{0.5cm}
\newline {H. Ohnishi, Y. Sada}
\address{Research Center for Electron Photon Science, Tohoku University, Sendai, Japan}
}
\maketitle
\begin{abstract}
\noindent The DA$\Phi$NE collider at INFN-LNF is a unique source of low-energy kaons, which was used by the DEAR, SIDDHARTA and AMADEUS collaborations for unique measurements of kaonic atoms and kaon-nuclei interactions. Presently, the SIDDHARTA-2 collaboration is underway to measure the kaonic deuterium exotic atom. With this document we outline a proposal for fundamental physics at the strangeness frontier for future measurements of kaonic atoms and kaon-nuclei interactions at DA$\Phi$NE, which is intended to stimulate discussions within the broad scientific community performing research directly or indirectly related to this field. 
\end{abstract}
  
\section{Introduction}\label{sec:4.1_Strangeness-TH}
\smallskip

\begin{figure}[htbp]
	\begin{center}
		\includegraphics[width=1.\linewidth]{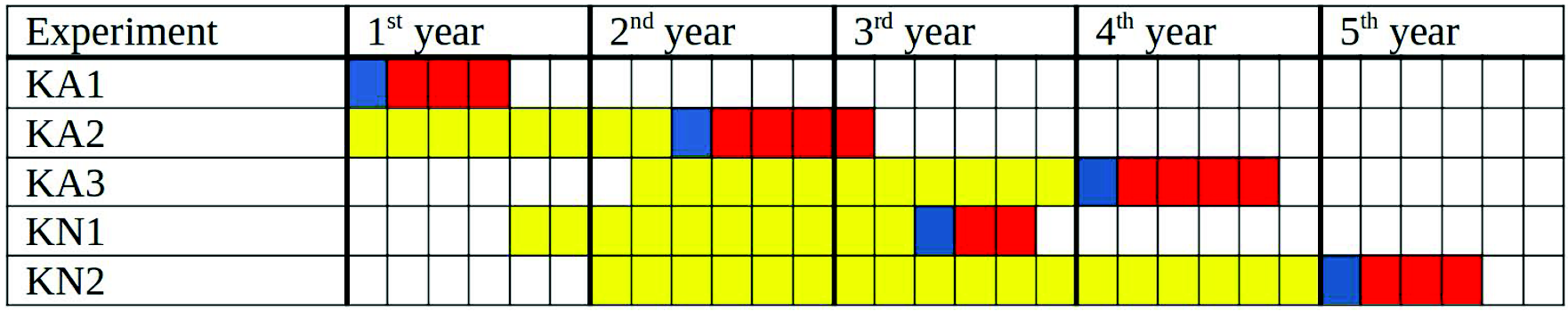}
	\hspace{0.2 cm}
    \caption{Schematic Gantt Chart for Fundamental physics at the Strangeness Frontier at the DA$\Phi$NE Proposal: KA1 (see Sec. \ref{ka1sec}), KA2 (see Sec. \ref{ka2sec}), KA3 (see Sec. \ref{ka3sec}), KN1 (see Sec. \ref{kn1sec}), KN2 (see Sec. \ref{kn2sec}). Yellow: preparation phase. Blue: installation phase. Red: data taking. }
		\label{fig0_hannes}
	\end{center}
	\vspace{-0.5cm}
\end{figure}

\noindent The DA$\Phi$NE  collider at INFN-LNF\cite{Zobov:2010,Milardi:2009} is a unique source of strangeness (kaons) in the world: it delivers low-momentum ($<$~140 MeV/c)  nearly monochromatic charged kaons,  generated by the decay of the $\phi$  resonance formed in electron-positron annihilation. This beam is  ideal for  experimental studies of low-energy kaon-nucleon/nuclei interactions.
\noindent  Huge advances in the development of fast spectroscopic  X-ray detector systems,  combined with the availability of the   DA$\Phi$NE  kaon beam, propelled an unprecedented progress in the field of strangeness studies with the DEAR\cite{Beer:2005} and the SIDDHARTA\cite{Bazzi:2011,Bazzi:2010,Bazzi:2009} experiments, which achieved the most precise measurement of kaonic hydrogen transitions to the ground level, and the first measurements of gaseous kaonic helium-3 and kaonic helium-4 transitions to the 2p level.
Presently, the SIDDHARTA-2 collaboration is  underway to perform the challenging kaonic deuterium measurement which, due to its difficulty (yield about ten times lower than for kaonic hydrogen and width at least twice larger), could not be performed till now. All these  measurements provide necessary inputs to and constraints on fundamental physics with strangeness, in particular to the theoretical models describing the strong interaction in the low-energy regime\cite{Curceanu:2020-sym,Gal:2016,Gal:2010,Horiuchi:2019}.
\noindent At the same time, the AMADEUS collaboration has performed unprecedented studies of kaon-nuclei interactions at low-energies\cite{Doce:2015,Piscicchia:2016,Piscicchia:2018,DelGrande:2018,DelGrande:2020}, delivering a wealth of new experimental data used for understanding low-energy QCD with strangeness, with important impact on astrophysics. 
DA$\Phi$NE confirmed to be  a unique machine in the world to perform  fundamental physics measurements  at the strangeness frontier.
\noindent A broad international community, growing on the backbone of the SIDDHARTA-2 collaboration,  proposes a strategy for future fundamental physics measurements at the strangeness frontier at   DA$\Phi$NE, which is supported by the EU STRONG-2020 Project (Grant Agreement n. 824093) both for the theoretical impact and  development, as well as for the realisation of new detector systems, and from various international and national agencies. 
This community proposes a five-year post-SIDDHARTA-2 scientific program for kaonic atoms and kaon-nuclei interactions  measurements of:

\begin{itemize}
\item[-]selected light and heavy kaonic atoms transitions (proposals KA1, KA2, KA3),
\item[-]low-energy kaon-nucleon scattering processes (proposal KN1),
\item[-]low-energy kaon-nuclei interactions (proposal KN2),
\end{itemize}

\noindent to be performed with a series of detector systems developed within the proponent community. These  measurements were selected as results of discussions and optimisations within a broad international community of theoreticians and experimentalists working in various fields, ranging from particle and nuclear physics to astrophysics and foundations of physics. A possible schematic schedule of the proposed measurements  is shown in the Gantt Chart  (see Fig. \ref{fig0_hannes}), and is being discussed in the following sections. 

\begin{figure}[htbp]
	\begin{center}
		\includegraphics[width=1.\linewidth]{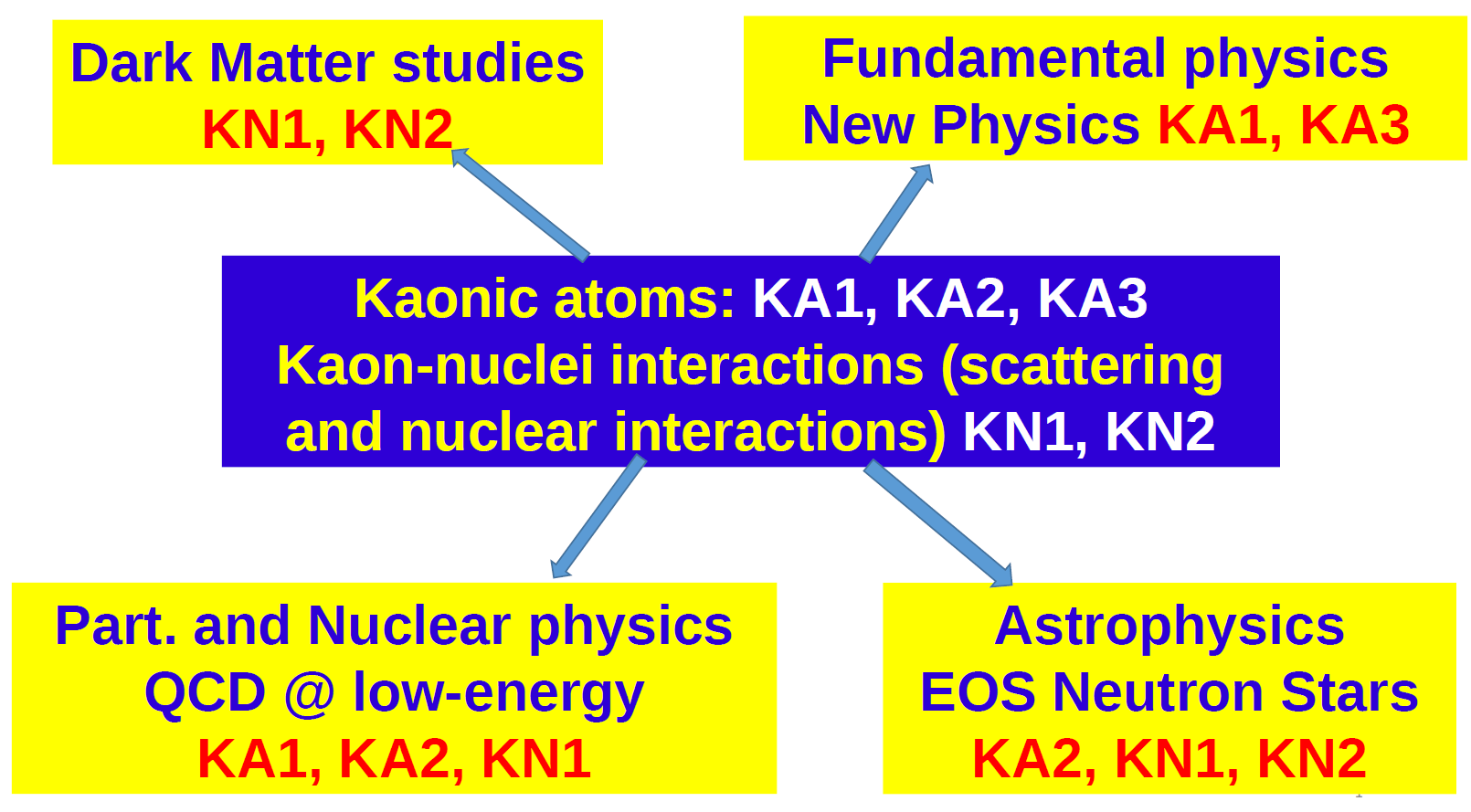}
	\hspace{0.2 cm}
    \caption{Impact of the Fundamental Physics at the Strangeness Frontier at DA$\Phi$NE studies in various sectors.}
		\label{fig1_catalina}
	\end{center}
	\vspace{-0.5cm}
\end{figure}

\noindent The frontier strangeness studies, which we propose to be performed  at DA$\Phi$NE in a strategic series of  measurements campaign, have a strong impact on many sectors of fundamental physics (see Fig. \ref{fig1_catalina}), among which:

\begin{itemize}
\item[-]\emph{Strong interaction:} QCD theory of the strong interaction in the low-energy regime,\cite{Curceanu:2020-sym,Gal:2016,Gal:2010,Horiuchi:2019};
\item[-]\emph{Astrophysics:} the Equation of State of Neutron stars\cite{Tolos:2017,Lonardoni:2013,Drago:2014,Ramos:2001,Djapo:2010,Logoteta:2019,Lonardoni:2015,Ribes:2019,Bonanno:2012};
\item[-]\emph{Dark Matter:} dark matter within the Standard Model, the Strange Dark Matter\cite{Merafina:2020};
\item[-]\emph{Fundamental physics:}  the measurement of the charged kaon mass and the solution of the ``kaon mass puzzle''\cite{PDG:2020,Bosnar:2020};
\item[-]\emph{Physics beyond the Standard Model (BSM):} measurements with extreme precision of kaonic atoms transitions, similar to the muonic atoms measurements, which have triggered the proton radius puzzle and the BSM searches with exotic atoms\cite{bsm:1,bsm:2}.
\end{itemize}

\noindent The proponents strongly believe that this is an opportunity which \emph{cannot be missed}, since we plan to measure fundamental interaction  processes  which could not be measured till now, and which have a big and concrete impact, in particle and nuclear physics, astrophysics, cosmology and foundational issues. 
\newline

\vspace{0.2cm}
\noindent \emph{ For details and recent discussions about the fundamental physics at the strangeness frontier, we refer to the dedicated workshop : ``Fundamental physics at the strangeness frontier at DA$\Phi$NE'', 25-26 February 2021, which can be found at:\vspace{0.2 cm} \newline https://agenda.infn.it/event/25725/overview.}
\newline

\vspace{0.2cm}
\noindent In Sec. \ref{sec:4.2_K-Atoms} we present proposals for kaonic atoms measurements, while in Sec. \ref{sec:4.3_K-Interactions} we outline ideas for kaon-nuclei interactions studies. Sec. \ref{sec:conclusions} completes the document with conclusions and perspectives.

\section{Fundamental physics with kaonic atoms}\label{sec:4.2_K-Atoms}
\smallskip

\noindent Since the first kaonic hydrogen measurement performed by the DEAR collaboration\cite{Beer:2005}, the   strangeness physics international collaboration at DA$\Phi$NE has sensibly grown, not only in terms of number of participants and involved institutions but, also, in terms of
expertise in technologies, such as cryogenic targets and detector systems, used  for performing low-energy strangeness nuclear physics studies, in particular  kaonic atoms measurements. 
The  strong impact of the  SIDDHARTA results, i.e., the  kaonic hydrogen and helium measurements\cite{Bazzi:2009,Bazzi:2010,Bazzi:2011}, and the  expected SIDDHARTA-2\cite{Curceanu:2013} kaonic deuterium one, together with the gathered experience and knowledge of our international community is  the driving force of the series of
future proposed experiments at the DA$\Phi$NE collider, having potential breakthrough impacts in several fields (see Sec. \ref{sec:4.1_Strangeness-TH}).

\par
\noindent We propose a series of kaonic atoms measurements, and briefly introduce  the  expected physics impact, the  kaonic transitions to be measured,
the required detector performances with a schematic setup and, finally, their feasibility and a first estimate of integrated luminosity requests.

\subsection{Selected heavy kaonic atoms measurements (KA1)}\label{ka1sec}
\vspace{0.3cm}

\begin{itemize}

\item \emph{Expected impact and physics motivation.} New precise measurements of heavy kaonic atoms transitions  have a strong  impact in two
main sectors shown in Sec. \ref{sec:4.1_Strangeness-TH}: the charged kaon mass puzzle (large uncertainty in the kaon mass and discrepancy between experiments)\cite{PDG:2020,FFF2021} and QCD at low-energy, by the investigation of 
in-medium effects related to the multi-nucleon interaction of the kaons. In this direction, high-Z targets can 
be exploited to measure transitions both to low and high {\it n} levels, where in the first case results about multi-nucleon interactions can be obtained
and in the second one, since high  {\it n} levels transitions are purely QED, the charged kaon mass problem could be addressed. The KPb atom is 
{\it  one of the two protagonists} of the charged kaon mass puzzle\cite{PDG:2020}. In all these experiments, the possibility to perform, for a single target, simultaneous measurements 
of atomic transitions from various {\it n} levels and with different $\Delta n$ will also help in the systematic errors minimisation and, in addition, 
provide useful information about the cascade processes in heavy kaonic atoms\cite{Koike:2005}.

\item \emph{Transitions to be measured.} We plan to perform a measurements campaign using  Se, Zr, Ta and Pb targets; the lines to be measured are,  
$\mathrm{KSe(6,5,4\rightarrow5,4,3)}$, $\mathrm{KZr(6,5,4\rightarrow5,4,3)}$, $\mathrm{KTa(8,7,6\rightarrow7,6,5)}$, $\mathrm{KPb(9,8,7,6\rightarrow8,7,6,5)}$.

\item \emph{Detector performances and setup.} The above mentioned transitions cover a wide energy range spanning from a few hundred keV up to a few MeV.
To measure with high precision these energies, we plan to implement a High Purity Germanium Detector (HPGe) which, thanks to a rate
capability up to $\mathrm{150\,kHz}$, could be positioned, with an appropriate shielding, close to the DA$\Phi$NE interaction point (IP) to maximise the geometrical efficiency.
This high rate capability comes from the usage of transistor reset preamplifier instead of conventional RC preamplifier and application of fast pulse digitizer for the processing of signals directly from preamplifier instead of conventional amplifier and ADC data processing chain.
To test the in-beam behaviour and assess the machine background effects on the system, an exploratory measurement is  planned to be
carried out in parallel with the SIDDHARTA-2 experiment. A sketch of the KA1 setup is shown in Fig. \ref{fig_geka_kahel_likam} upper panel, where the active part of the HPGe detector (e), the target (c) 
and its holder (d) are shown, together with part of the mechanical support and the lead shielding (f), the SIDDHARTA-2 luminosity monitor\cite{Skurzok:2020} to be used as trigger (b) and
the DA$\Phi$NE beam pipe (a)\cite{Bosnar:2020}.

\begin{figure}[htbp]
	\begin{center}
		\includegraphics[width=0.6\linewidth]{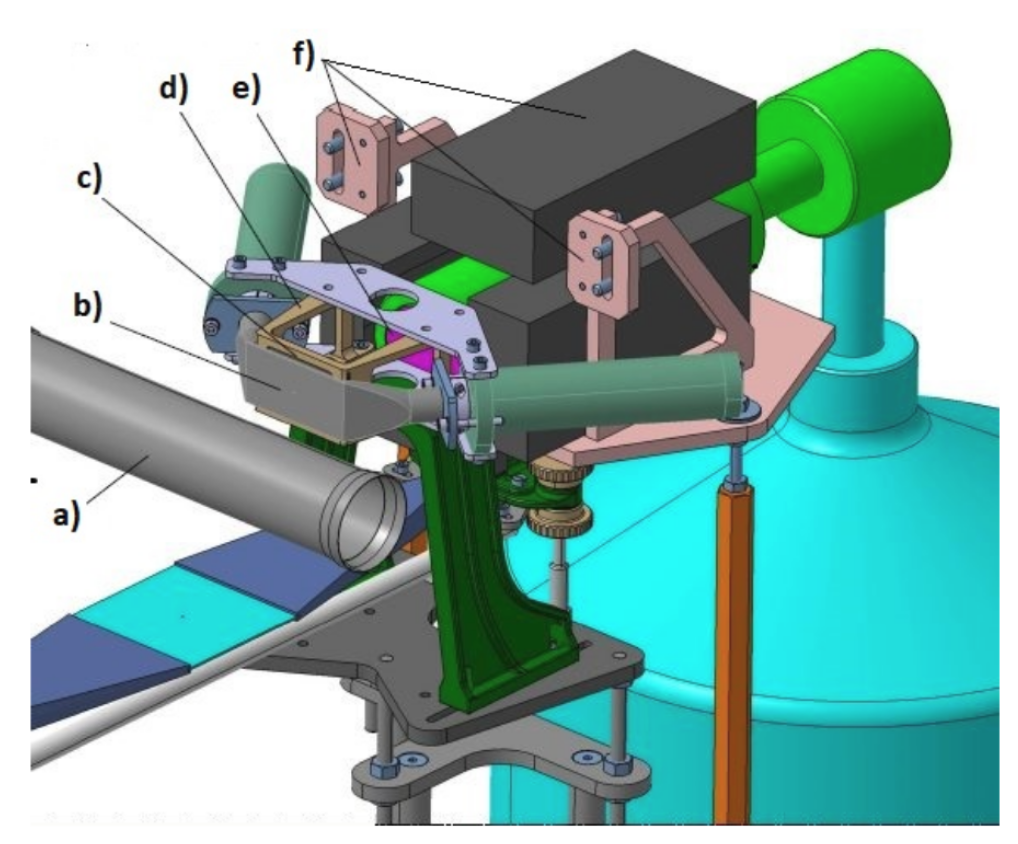}
		\includegraphics[width=0.6\linewidth]{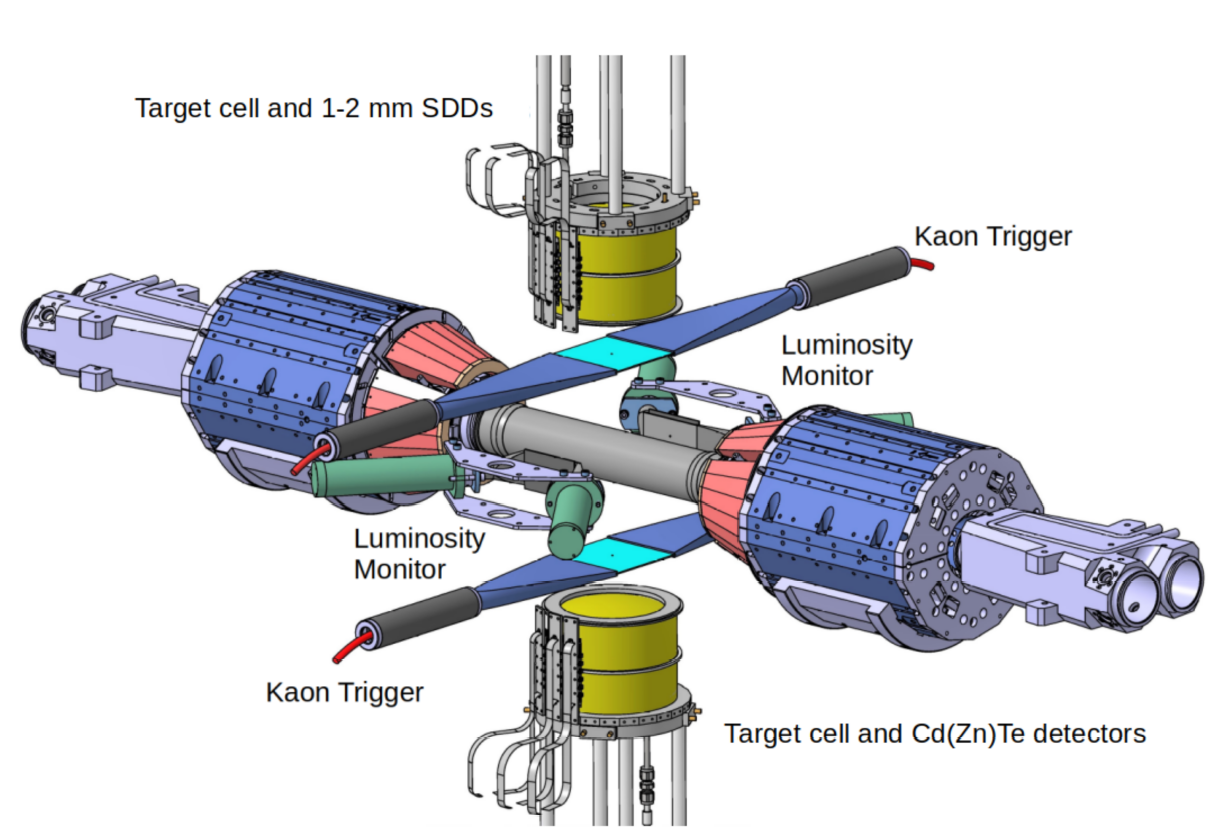}
		\caption{Setups of the KA1 (up) and KA2 (down) experiments (see text for details).}
		\label{fig_geka_kahel_likam}
	\end{center}
\end{figure}

\item \emph{Luminosity estimate and feasibility.} The HPGe detector and its readout system have been already assembled at the University of Zagreb, where several tests confirmed its excellent performances in terms of energy resolution 
($FWHM \simeq \, 1,06\, \mathrm{keV}\, @\, 302.9\,\mathrm{keV}$ and $1,67\, \mathrm{keV}\, @\, 1,33\, \mathrm{MeV}$) and timing\cite{Bosnar:2020}. 
Assuming SIDDHARTA-like conditions ($\mathcal{L}_{int}\simeq10\,\mathrm{pb^{-1}/day}$), from MC simulations we obtain that measurements with a
precision of few eV could be achieved with a delivered integrated luminosity of $\mathrm{360\,pb^{-1}}$ considering {\it the efficiency for each target}.
These results can be obtained in an overall 8 months experiment: 2 months for the setup mounting and optimisation and 6 months of data taking running with all four targets.

\end{itemize}

\subsection{Selected light kaonic atoms measurements (KA2)}\label{ka2sec}
\vspace{0.3cm}

\begin{itemize}

\item \emph{Expected impact and physics motivation.} Measurements of light kaonic atoms transitions are fundamental to address important issues like kaon-nuclei potential and chiral models below threshold and the nature of the ambiguous $\Lambda(1405)$. These results  are important also in astrophysics: search for dark matter with strangeness and the equation of state for neutrons stars\cite{Tolos:2017,Lonardoni:2013,Drago:2014,Ramos:2001,Djapo:2010,Logoteta:2019,Lonardoni:2015,Ribes:2019,Bonanno:2012}.
In particular, the first ever measurement of the $\mathrm{K^{3,4}He(2p\rightarrow1s)}$ transition, if possible, could  put stronger constraints on the theoretical models describing the kaon-nucleon interaction in systems with more than two nucleons\cite{FFF2021,GalFri:2020pc}.
Information on the nature of the $\Lambda(1405)$  state can be obtained from the upper level transitions of light kaonic atoms, like different isotopes of KLi, KBe and KB\cite{Wycech:2020vpl}, (see Sec. \ref{sec:4.1_Strangeness-TH}).
	
\item \emph{Transitions to be measured.} We plan to use $\mathrm{^{3,4}He,^{6,7}Li,^{9}Be,^{10,11}B}$ targets to perform measurements of both low level and high level transitions with $\Delta n=1,2,...5$,  and energies in the range 10-100 keV.

\item \emph{Detector performances and setup.} We plan to perform the above mentioned measurements using two SIDDHARTA-like setups in parallel, to be placed above and below
the DA$\Phi$NE IP. The first one, devoted to the lowest energy region, will make use of $1\,\mathrm{mm}$ thick Silicon Drift Detectors (SDD) having large efficiencies
in the 10-40 keV range; the second one will concentrate on the 40-100 keV lines by means of Cd(Zn)Te devices. $1\,\mathrm{mm}$ thick SDD detectors are being built, while R\&D CdTe is ongoing in the framework of the Work-Package JRA8-ASTRA of  the EU STRONG-2020 project, in which our community is involved\cite{ASTRA}. Both  detectors have the required energy resolutions (FWHM $\simeq 150-200\,\mathrm{eV} (SDD)\, / \,0,5-1\,\mathrm{keV}$ (Cd(Zn)Te)) to clearly separate the lines of interest.  
A simple schematic of the possible setup is shown in Fig. \ref{fig_geka_kahel_likam} lower panel, where the two replica of the target cell, (partially) surrounded by the holders for the detectors,
are shown above and below  the IP; two pairs of scintillators, to be used as Kaon Trigger and Luminosity Monitor like in the SIDDHARTA-2 experiment, are shown as well.

\item \emph{Luminosity estimate and feasibility.} The feasibility of both experiments is based on two solid starting points. 
First, the production of $1\,\mathrm{mm}$ thick SDDs has already been financed by the INFN CSN3; in addition, our collaboration already possesses a 
strong knowledge of their readout electronics, handling and calibration procedure, gathered  during the SIDDHARTA and SIDDHARTA-2 experiments.
Second, prototypes of Cd(Zn)Te detectors and their related readout electronics will be  delivered by the JRA8-ASTRA.
Assuming SIDDHARTA-like conditions we estimate a required integrated luminosity of $800\,\mathrm{pb^{-1}}$ for kaonic helium and of $400\,\mathrm{pb^{-1}}$ for the other targets.
This results in an overall 26 months time
schedule: 12 months for the design and production of detectors, electronics and mechanical supports,
6 months for the setup calibration, test, mounting and optimisation, and 8 months of data taking.

\end{itemize}

\subsection{ Ultra-high precision measurements of selected exotic atoms (KA3)}\label{ka3sec}
\vspace{0.3cm}

\begin{itemize}

\item \emph{Expected impact and physics motivations.} A new exciting possibility in the investigation of kaonic atoms is represented by extreme (sub-eV) precision measurements of their  transition energies. 
For example, a particularly significant measurement would be that of the relative difference in the $\mathrm{3d\rightarrow2p}$ transition between $\mathrm{K^3He}$ and $\mathrm{K^4He}$ (known as the \emph{isotopic shift}) and of their widths, 
whose measurement with a precision below 1 eV might represent a breakthrough in the field.
For a proper determination of this shift, it is as well necessary to have an accurate determination of the electromagnetic values of the 2p energy levels which,
in turn, are influenced by the precision of the $K^-$ mass value, which could be also extracted with sub-eV precision measurements of kaonic atoms. 
Moreover, it turns out that the systematic uncertainty  of the $D^0$-meson mass is also limited by the precision on the charged kaon
mass\cite{PhysRevD.88.071104}. Also, the masses of excited charm mesons, whose direct measurements
are rather uncertain, e.g. those of $D_1(2420)^0$, $D_2^{*}(2460)^0$ and $D_{s1}(2536)^{\pm}$, are determined from the fit based on high-precision measurements of mass and mass difference of the $D^0$, $D^{\pm}$ and $D^{\pm}_s$ states\cite{PDG:2020}. 
The precision on the $D^0$ mass also impacts on the mixing parameters in the $D^0-\overline{D}^0$ system\cite{PhysRevD.88.071104}.
In the long run, a more accurate kaon mass determination may become appealing for first-principle calculations on
the lattice\cite{refId0}.
Finally, measurements of very narrow kaonic atoms lines (few eV or below eV FWHM) in transitions from upper levels, from which important information on the $\Lambda(1405)$ nature can be extracted\cite{Wycech:2020vpl}, could only be performed with such a 
high resolution experiment.

\item \emph{Transitions to be measured.} To address the scientific questions mentioned above and in Sec. \ref{sec:4.1_Strangeness-TH}, we plan to perform simultaneous measurements of 
$\mathrm{K^{3,4}He(7,6,5,4,3\rightarrow2)}$ as well as $\mathrm{KN(12,11,10,9,8,7,6\rightarrow7,6,5)}$ transitions, where the first ones will give a measurement of the isotopic shift and  
the second ones represent inputs to the kaon mass precise determination\cite{PDG:2020,Bosnar:2020}, while from both a rich data set for kaon cascade processes can be retrieved\cite{Koike:2005}. 
Measurements of light-Z kaonic atoms with very narrow upper level transitions\cite{Wycech:2020vpl}, are also foreseen.
All the lines of interest fall in the 6-15 keV range. 

\item \emph{Detector performances and setup.} To achieve the required challenging precisions, detectors with  $FWHM\simeq\,5-10\,\mathrm{eV}\,@\,6-10\,\mathrm{keV}$ resolution, namely more than 
one order of magnitude better than the standard large area solid state devices, are required. We plan to exploit the 
resolution and efficiency performances of the Highly Annealed Pyrolitic Graphite (HAPG) mosaic crystals based Von Hamos spectrometer, developed by the VOXES collaboration
at LNF\cite{VOXES:2020,VOXES:2019,VOXES:2018,VOXES:2017}. A drawing of the proposed setup, consisting of 8 spectrometer arms devoted to a specific energy range each, is shown in Fig. \ref{fig_numes},
while the present setup of the VOXES spectrometer, which represents a single arm, is shown in Fig.\ref{fig_voxes}.
Kaons emitted from the IP are first detected by a scintillator and SiPMs based trigger system, before entering in a cylindrical gaseous target cell surrounding the beam pipe,
where they form the kaonic atoms; the X-ray photons emitted during the transitions are then measured by a specific energy range spectrometer having a resolution of few eV.

\begin{figure}[htbp]
	\begin{center}
		\includegraphics[width=.9\linewidth]{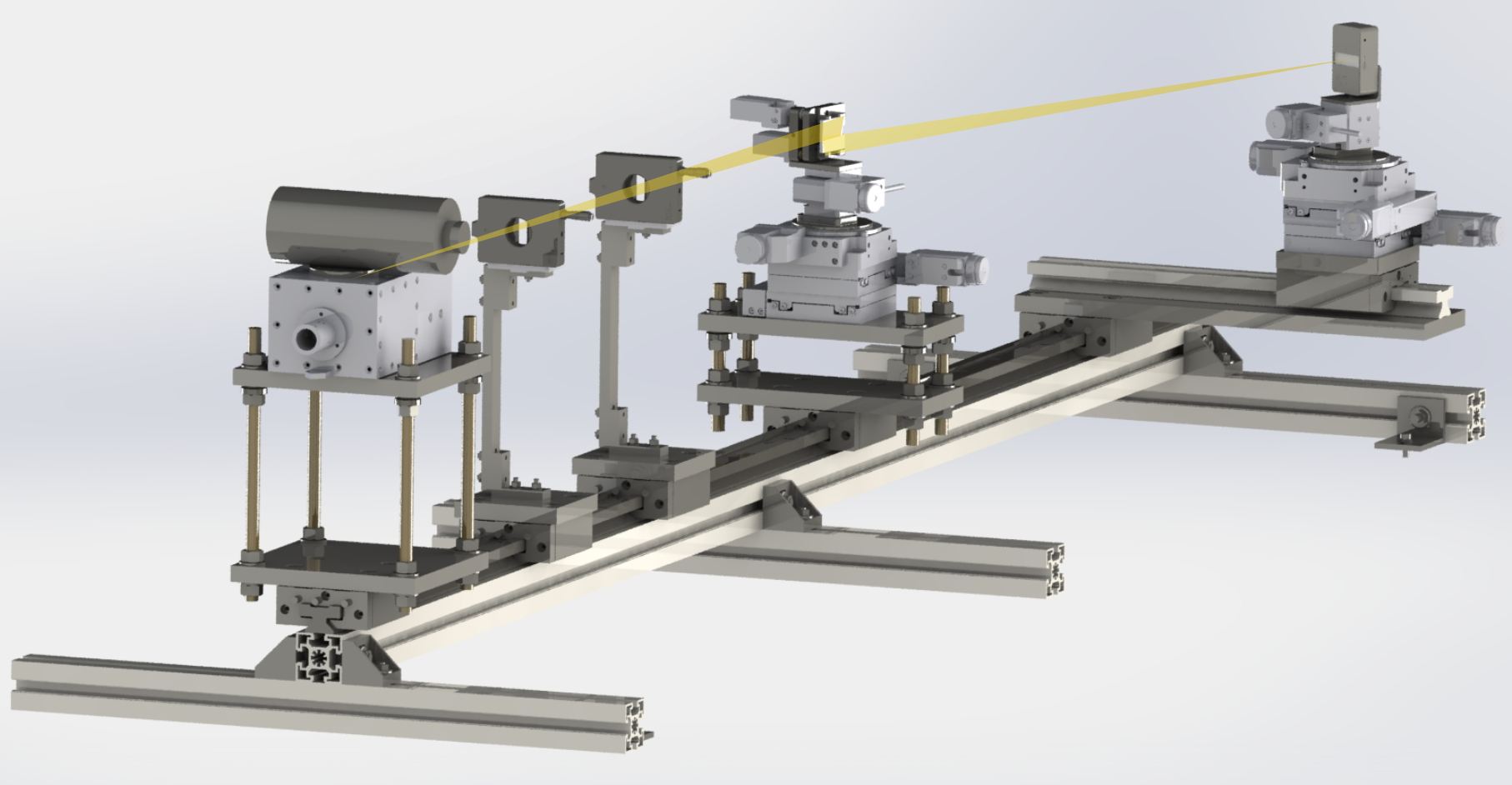}
		\caption{Sketch of the present VOXES setup\cite{VOXES:2020}.}
		\label{fig_voxes}
	\end{center}
\end{figure}

\begin{figure}[htbp]
	\begin{center}
		\includegraphics[width=0.9\linewidth]{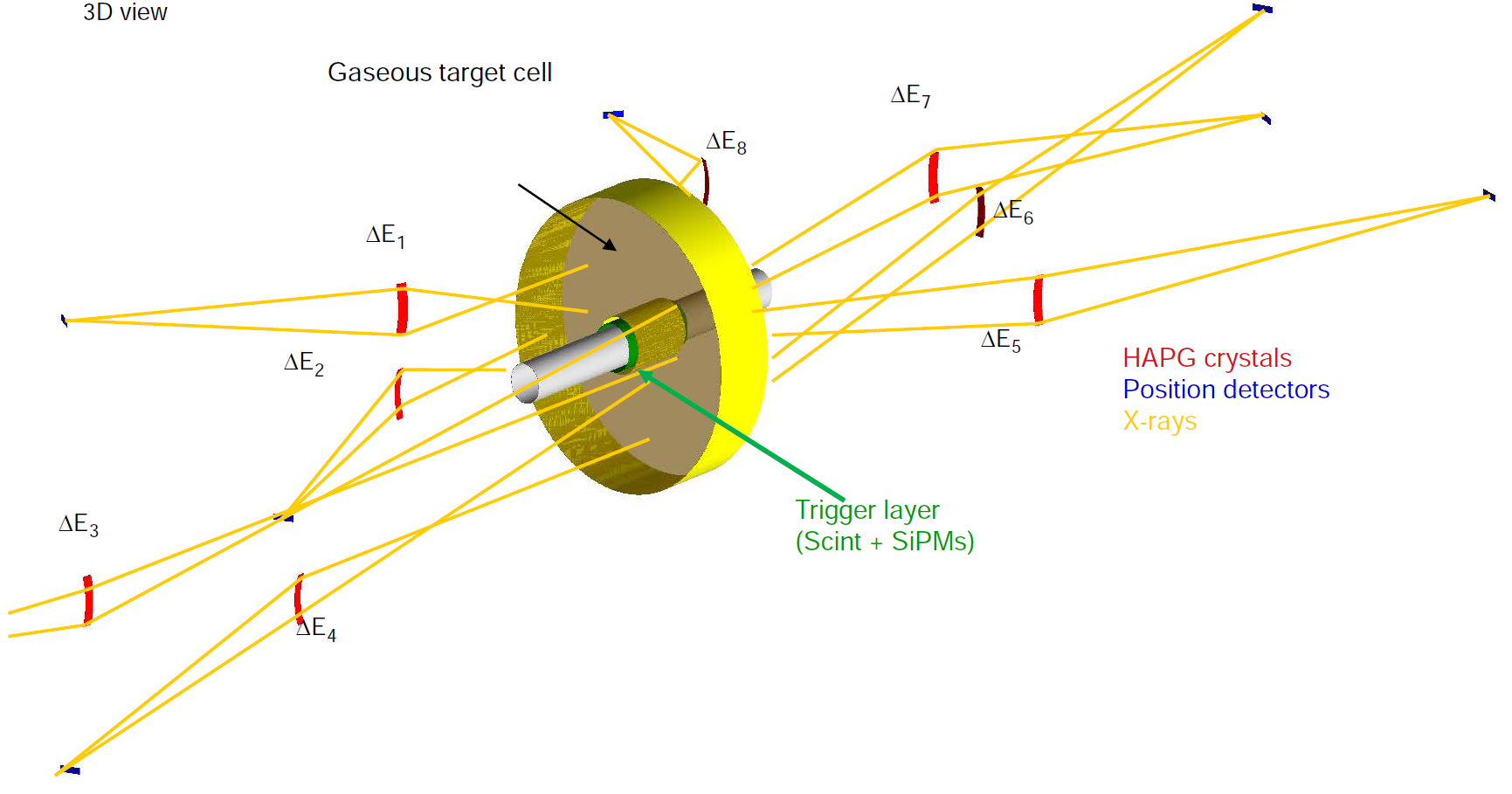}
		\caption{Sketch of a possible setup of the KA3 experiment (see text for details).}
		\label{fig_numes}
	\end{center}
	\vspace{-0.5cm}
\end{figure}

\item \emph{Luminosity request and feasibility.} The foreseen performances of the KA3 experiment are supported by very encouraging preliminary data obtained within the VOXES project\cite{VOXES:2020,VOXES:2019,VOXES:2018,VOXES:2017}. As an example, results obtained with targets of Fe, Cu, Ni, Zn, Mo and Nb are shown in Fig. \ref{fig_voxres}; in the upper pads, the extremely low resolutions and the resulting high precisions achieved with single element targets (Fe and Cu) are shown,
while in the lower ones the possibility to have spectra with larger dynamic range without dramatic worsening of the resolutions (left) and spectra in the 15-20 keV range with few eV precisions (right) is also demonstrated. 
In all the pads of Fig.\ref{fig_voxres}, the effective source size $S_0'$ for each measurement is reported, representing the real breakthrough of the VOXES spectrometer (see \cite{VOXES:2020} for more details).
Consistent ray-tracing simulations have been performed, allowing a successful design of the setup. The in-beam behaviour of the crystals and the position detectors is planned to be assessed in a  run in parallel with SIDDHARTA-2. To fulfill the scientific program of KA3, $\simeq 2000\,\mathrm{pb^{-1}}$ of delivered luminosity are estimated.
The estimated time duration of the whole experiment will be assessed in 32 months: 16 months devoted to crystals, detectors and mechanics design and production, 6 months to test and calibrations, 2 months for mounting and, finally, the last 8 months to the data taking.   
\begin{figure}[htbp]
	\begin{center}
		\includegraphics[width=1.\linewidth]{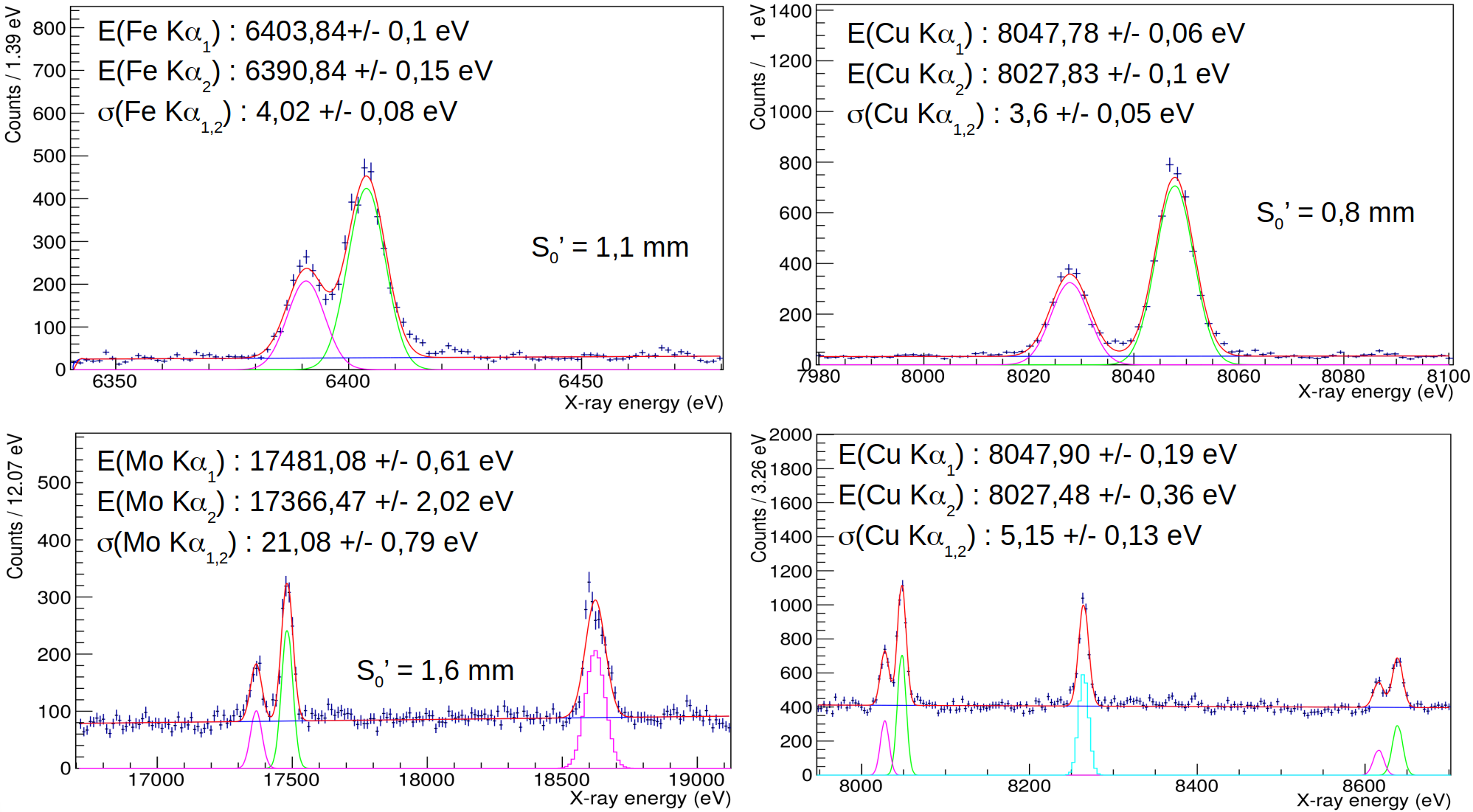}
		\caption{Example of precisions and resolutions obtained by VOXES\cite{VOXES:2020}.}
		\label{fig_voxres}
	\end{center}
	\vspace{-0.5cm}
\end{figure}
\end{itemize}

\section{Kaon-nuclei interaction studies at the low-energy frontier}\label{sec:4.3_K-Interactions}
\smallskip

\subsection{Expected impact and physics motivation}

\begin{figure}[htbp]
	\begin{center}
		\includegraphics[width=1.\linewidth]{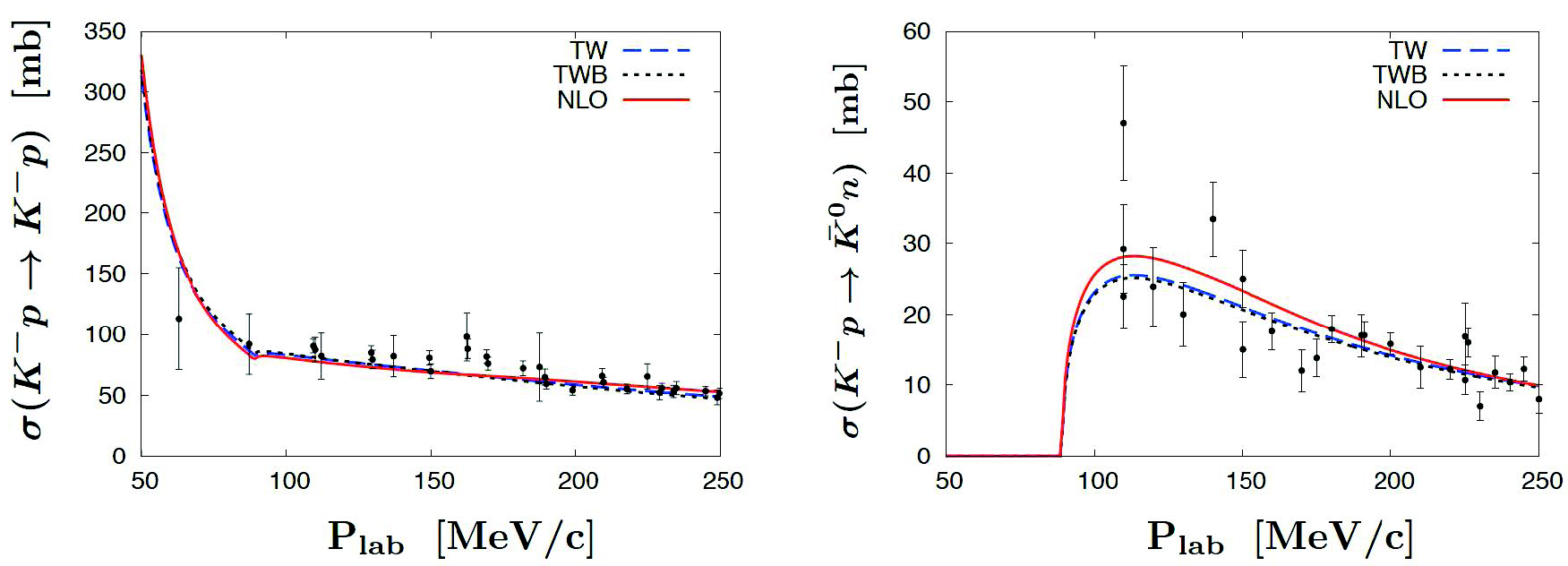}
	\hspace{0.2 cm}
    \caption{Two examples from\cite{Ikeda:2012}: (left) experimental data set for the elastic kaon-proton reaction, (right) the  $K^{-}p \rightarrow \overline{K}^0 n$ inelastic channel, both fitted with theoretical models. Data below 100 MeV/c have large error bars or are missing. }
		\label{fig1_hannes}
	\end{center}
	\vspace{-0.5cm}
\end{figure}

\noindent The physics of the strong interaction manifests in the dynamics driven by the chiral symmetry breaking of low energy QCD. Explicit symmetry breaking by the mass of the strange quark plays an important role when extending the framework to flavour SU(3). Low-energy interactions of antikaons with nucleons and nuclei, measured with the highest possible precision, are excellent probes to explore and improve the understanding of this fundamental
topic - the interplay between spontaneous and explicit chiral symmetries breaking. 
High-precision $\mathrm{K^-p}$ threshold data, derived from accurate results for kaonic hydrogen from DEAR\cite{Beer:2005} and recently from SIDDHARTA\cite{Bazzi:2011}, set important constraints for theoretical approaches.  
Precise total and differential cross sections of low-energy kaon-nucleon reactions are however missing. The available experimental data below 100 MeV/c are very scarce and with large errors bars, see for example Fig. \ref{fig1_hannes}.
\noindent Kaon induced $\Lambda$(1405)  production on light nuclear targets can provide crucial information to solve the longstanding debate on the nature of the $\Lambda$(1405) state. In particular, the process $K^{-}d \rightarrow \Sigma^{0} \pi^{0} n$ , for a  $K^{-}$ momentum of the order of 120 MeV/c, was investigated in the context of both phenomenological potential approaches\cite{Esmaili:2009} and chiral unitary models\cite{Jido:2010}  and was found to be the golden channel to unveil the possible existence of a high mass (1420 MeV/c$^{2}$) pole of the $\Lambda$(1405), and hence to (definitely) solve this fundamental problem.

\subsection{Kaon-nucleon elastic scattering KN1}\label{kn1sec}

\noindent The goal of the KN1 experiment  is the measurement of the low-energy scattering process of kaons, and of the $\Lambda$(1405) kaon induced production,  on various targets such as hydrogen, deuterium, helium-3 and helium-4. The measurements of  particles with such low momenta represent a big experimental challenge. The component of the experimental apparatus will be an active Time Projection Chamber (TPC) with a Gas Electron Multiplier (GEM) readout device, which  allows the study of kaons interactions directly in the TPC gas/target volume, without additional material. We have already started to develop, in the framework of the EU programmes HadronPhysics3 and  STRONG-2020, active target concepts using GEM-TPCs, which are well suited for this type of studies.
 The experimental setup is made of three detector parts: the kaon monitor, 
the active target detector (a GEM-based TPC) and  the charged kaon detector.

\begin{itemize}

  \item \emph{The kaon monitor}  consists of plastic scintillator pads (see Fig. \ref{fig2_hannes} top), which are read out on both sides with Silicon Photo-Multipliers (SiPMs). The scintillator pads are arranged around the interaction region, covering the TPC kaon entrance window. The kaon monitor detects the  charged kaon-pairs produced via the $\phi$-decay, which are emitted back to back. This topology allows to set up a clear trigger scheme, with excellent timing (starting point) and in addition allows to reduce background events.

\begin{figure}[htbp]
	\begin{center}
		\includegraphics[width=0.62\linewidth]{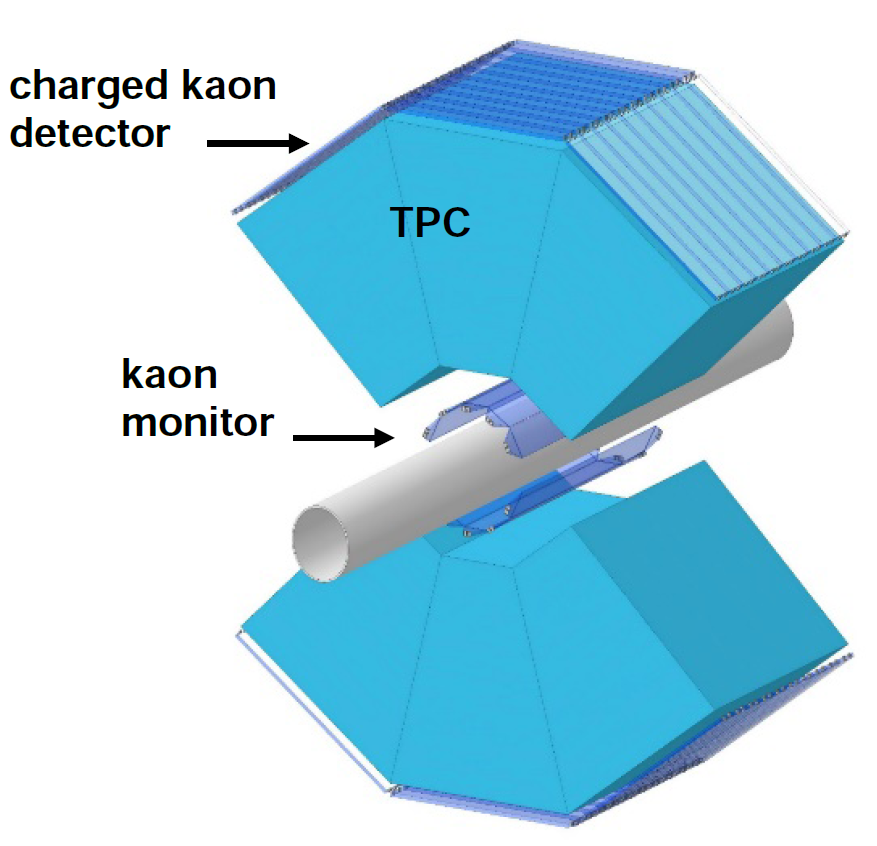}
	\hspace{0.2 cm}
		\includegraphics[width=1.\linewidth]{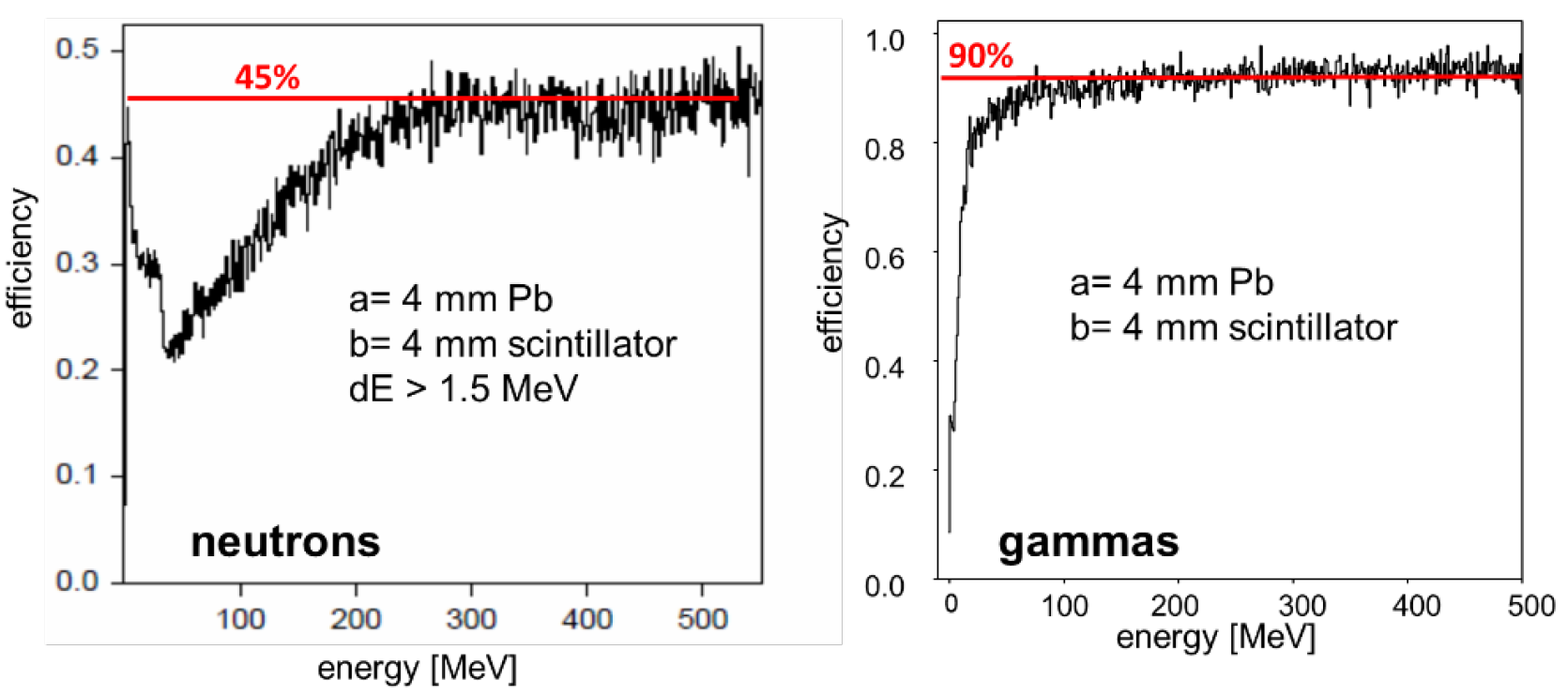}
		\caption{Sketch of the possible setup for the KN-elastic, KN1, experiment (top) and efficiency simulation with Geant4 of the detector for neutrons and gammas (bottom).}
		\label{fig2_hannes}
	\end{center}
	\vspace{-0.5cm}
\end{figure}

 \item \emph{The active GEM-TPC} as a low mass target and detector, motivated by the need to study the low energy interactions of kaons with nuclei in a complete way, using an almost massless  system.
 This detection technique requires the use of low-radiation length materials and very pure light gases such as hydrogen, deuterium, helium-3, helium-4, etc. To evaluate the GEM-TPC performances, a 10 $\times$   10 cm$^{2}$ prototype with a drift gap of 15 cm has been realised. The detector was tested at the $\pi$M1 beam facility of the Paul Scherrer Institut (PSI) with low momentum pions and protons\cite{PoliLener:2015}. In addition, tests with pure hydrogen and helium gases without impurities were tested with a triple GEM device at SMI in Vienna in the framework of a diploma thesis\cite{master:2015}. A picture of the GEM-TPC developed at INFN is shown in Fig.\ref{fig_gem-tpc}

\begin{figure}[htbp]
	\begin{center}
		\includegraphics[width=0.7\linewidth]{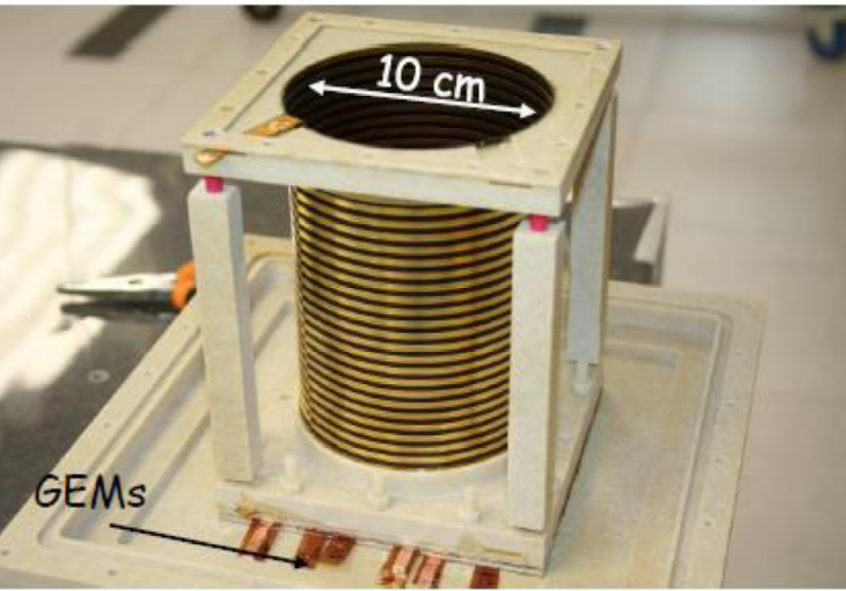}
		\caption{Prototype of an active GEM-TPC developed at LNF\cite{PoliLener:2015}.}
		\label{fig_gem-tpc}
	\end{center}
	\vspace{-0.5cm}
\end{figure}

 \item \emph{ The charged kaon detector}: made of plastic scintillator pads,  read out with SiPMs, covering the outer part of the TPC window completely. The scintillator pads have to be thick enough to stop all kaons, with the goal to distinguish between negative and positive charged kaons (see  Fig. \ref{fig2_hannes}).
 \end{itemize}.  
 
\noindent The layout of the KN1 detector system is in  an advanced state. The kaon monitor will be quite similar to that already used for SIDDHARTA-2, while the charged kaon detector will use the same components as the Veto-2 detector system of SIDDHARTA-2. The design studies for the GEM-TPC used as active target is almost ready and a first prototype should be built within this year in collaboration with colleagues from ELPH, Tohoku University.

\subsection{Kaon-nuclei interactions (KN2)}\label{kn2sec}

\noindent To study inelastic channels, in particular the process:  $K^{-}d \rightarrow  \Lambda(1405)  n  \rightarrow  \Sigma^{0}  \pi^{0} n$,  it is necessary to detect neutrons, as well as gammas.  The resonant formation of the high mass  $\Lambda(1405)$ pole, is predicted in\cite{Jido:2010} to be enhanced by selecting forward neutrons, with respect to the incident $K^{-}$, in the center-of-mass frame. Therefore, a new detector concept to detect neutral particles is under study. The idea is to use lead plates separated by liquid scintillator. The scintillation light will be picked-up by wavelengths shifting fibres, which are led through on both sides  of the detector volume and read-out on both sides by SiPMs. Monte Carlo studies (Geant4) have shown (see Fig. \ref{fig2_hannes} right) that such a design is quite suitable for our purpose. In detail, one detector module could have a height of 20 cm and a thickness of 25 cm with a length of 80 cm and consists of 32 layers of lead (4 mm thick) with 4 mm gap between lead layers filled with liquid scintillator. The expected efficiency for neutrons is 45$\%$, while for gammas is 90 $\%$.
The detector setup will use the same components as for the KN1 experiment and in addition 20 modules of the lead-scintillator sandwich detector will be placed around the KN1 setup. \newline 

\noindent We are also studying the feasibility of selected hypernuclei measurements required by theoreticians, by adding HPGe detectors (see KA1) to the setup; this suggestion came out from the discussion during the Strangeness workshop:\vspace{0.2 cm}\newline \emph{https://agenda.infn.it/event/25725/overview} \vspace{0.2 cm}\newline of 25-26 February 2021.

\subsection{Time schedule}

\noindent For the KN1 experiment preliminary Monte Carlo simulations have been performed with Geant4 showing that with an integrated luminosity of 10 pb$^{-1}$  per day we expect $\sim$ 2000 $K^{-}$p elastic scattering events within about 30 days. For the proposed programme to measure hydrogen, deuterium, helium-3 and helium-4 a total of four months of beam time is necessary. \vspace{0.2 cm}

\noindent For the KN2 experiment Monte Carlo simulations are ongoing. The typical measuring time for one target is of the order of 5-6 weeks with an integrated luminosity of 10 pb$^{-1}$ per day.

\section{Conclusions and perspectives}\label{sec:conclusions}

\noindent In this document we have outlined a proposal for future measurements dedicated to the studies of fundamental physics at the strangeness frontier using the unique kaon beam delivered by the $DA\Phi$NE Collider at LNF-INFN. We briefly introduced the scientific motivations, together with the series of kaonic atoms and kaon-nuclei interactions that we propose to measure with dedicated setups, in a 5 years planning, to start when the present SIDDHARTA-2 experiment completes its data taking campaign.
The proposed measurements have the potential to guide the developments of physics at the strangeness frontier in the coming decade, being fundamental not only for a better understanding of the strong interaction (QCD), but also in astrophysics and particle physics.
The aim of this document is to present to the broad scientific community, performing research directly or indirectly connected to strangeness studies, our plans and to stimulate a discussion to optimise/integrate our strategy, and to gather together all those interested to participate to this new
adventure in strangeness physics.

\section*{Acknowledgement}

\noindent We want to acknowledge the LNF Director F. Bossi, P. Gianotti, E. Nardi and the DA$\Phi$NE team. 
Part of this work was supported by the Croatian Science Foundation under the project 8570, the Austrian Science Fund (FWF), P24756-N20; the Austrian Federal Ministry of Science and Research (BMBWK), 650962/0001 VI/2/2009; the Ministero degli Affari Esteri e della Cooperazione Internazionale, Direzione Generale per la Promozione del Sistema Paese (MAECI); the Strange Matter Project, the Polish Ministry of Science and Higher Education through grant No. 7150/E-338/M/2018; the Ministry of Science and Higher Education of the Russian Federation, Agreement 14.W03.31.0026 and the EU STRONG-2020 project (grant agreement No. 824093).

\bibliography{bib4_Strangeness}{}
\end{document}